# scientific reports

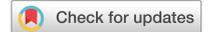

OPEN

# Fully fluorinated non-carbon compounds $NF_3$ and $SF_6$ as ideal technosignature gases

Sara Seager[1,2,3✉], Janusz J. Petkowski[1,4], Jingcheng Huang[1], Zhuchang Zhan[1], Sai Ravela[1] & William Bains[1,5]

Waste gas products from technological civilizations may accumulate in an exoplanet atmosphere to detectable levels. We propose nitrogen trifluoride ($NF_3$) and sulfur hexafluoride ($SF_6$) as ideal technosignature gases. Earth life avoids producing or using any N–F or S–F bond-containing molecules and makes no fully fluorinated molecules with any element. $NF_3$ and $SF_6$ may be universal technosignatures owing to their special industrial properties, which unlike biosignature gases, are not species-dependent. Other key relevant qualities of $NF_3$ and $SF_6$ are: their extremely low water solubility, unique spectral features, and long atmospheric lifetimes. $NF_3$ has no non-human sources and was absent from Earth's pre-industrial atmosphere. $SF_6$ is released in only tiny amounts from fluorine-containing minerals, and is likely produced in only trivial amounts by volcanic eruptions. We propose a strategy to rule out $SF_6$'s abiotic source by simultaneous observations of $SiF_4$, which is released by volcanoes in an order of magnitude higher abundance than $SF_6$. Other fully fluorinated human-made molecules are of interest, but their chemical and spectral properties are unavailable. We summarize why life on Earth—and perhaps life elsewhere—avoids using F. We caution, however, that we cannot definitively disentangle an alien biochemistry byproduct from a technosignature gas.

The search for signs of life beyond Earth is an increasingly popular scientific research area as telescope capability advances. For exoplanets, the successful launch and operation of the James Webb Space Telescope (JWST)[1], brings in a new level of precision for exoplanet atmosphere measurements, fostering hope that the community may find signs of life through the detection of exoplanet atmosphere biosignature gases within a decade[2–8]. Yet, as the community pushes deeper into biosignature gas identification, a dawning conclusion is that biosignature gases will always be stymied by false positives, fatal to robust conclusions about the presence or absence of life. Because of this, there is a growing interest in the study of technosignatures as indicators of life beyond Earth.

Technosignatures are astronomically detectable signs of a technological society on an exoplanet[9–14]. The concept of technosignatures dates back decades and includes several categories, such as: Lowell's canals on Mars[15]; artificial illumination of a planet (e.g.,[16,17]; megastructures (e.g.,[18,19]); waste heat (e.g.,[18,20,21]); and artificial (i.e., non-terrestrial) artifacts (e.g.,[22–24]). Technosignature research has recently been growing, e.g.,[11,12,25,26].

Technosignature gases, as a subset of technosignatures in general, are artificially produced gases that can accumulate to detectable levels in an exoplanet atmosphere. A technosignature gas can either be emitted for a specific purpose or released as a by-product of industrial civilization[12,16,27]. Proposed technosignature gases center around industrial pollutants such as chlorofluorocarbons (CFCs), hydrofluorocarbons (HFCs), and perfluorochemicals (PFCs)[12,28]. A conclusion from the above references is that detection of CFCs in exoplanet atmospheres, under the most favorable assumptions, would require 100–500 h of JWST in-transit observation time (Table 1). This is increased to 200–1000 h of total observing time, when taking into account an out-of-transit equivalent baseline time. This can be compared to the current JWST time allocation to individual exoplanets which are typically up to 20 h, with a few planet full phase curves allocated around 40–50 h, and one transiting exoplanet atmosphere exception at 70 h[29]. The exception to the long-required observation times is for the hypothetical and highly favorable case of a terrestrial-size planet transiting a tiny star—a white dwarf star which itself is about the size of Earth and is a dead remnant of a Sun-like star. This scenario would require only tens of hours of in-transit time[28,30,31]. See the SI for more details on CFCs and other suggested technosignature gases.

[1]Department of Earth, Atmospheric and Planetary Sciences, Massachusetts Institute of Technology, 77 Massachusetts Avenue, Cambridge, MA 02139, USA. [2]Department of Physics, Massachusetts Institute of Technology, 77 Massachusetts Avenue, Cambridge, MA 02139, USA. [3]Department of Aeronautics and Astronautics, Massachusetts Institute of Technology, 77 Massachusetts Avenue, Cambridge, MA 02139, USA. [4]JJ Scientific, 02-792 Warsaw, Poland. [5]School of Physics and Astronomy, Cardiff University, 4 The Parade, Cardiff CF24 3AA, UK. ✉email: seager@mit.edu

| Technosignature gas | Planet type | Atmosphere type | Star type | JWST detectable abundance | Observation time |
|---|---|---|---|---|---|
| CFC 14 ($CF_4$) | Rocky-Earth-size | $N_2$-dominated | White Dwarf | 10× present Earth's atm. abundance | 40.8 h[28] |
| CFC-11 ($CCl_3F$) | Rocky-Earth-size | $N_2$-dominated | White Dwarf | 10× present Earth's atm. abundance | 28.8 h[28] |
| CFC-11 (CCl3F) | Rocky-Earth-size | $N_2$-dominated | M-dwarf | 0.225 ppb | 100–300 h[12] |
| CFC-12 (CCl2F2) | Rocky-Earth-size | $N_2$-dominated | M-dwarf | 0.515 ppb | 100–300 h[12] |
| $NO_2$ | Rocky-Earth-size | $N_2$-dominated | M-dwarf | 20× present Earth's atm. abundance | *500 h[32] |
| $NH_3$, $N_2O$ | Rocky-Earth-size | $N_2$-dominated | G-dwarf | N/A | N/A[33] |

**Table 1.** Summary of proposed exoplanet atmosphere technosignature gases including simulated in-transit observation times. *not explicitly stated whether this is in-transit or both in-transit and out-of-transit time. N/A means no information available.

Here we propose nitrogen trifluoride ($NF_3$) and sulfur hexafluoride ($SF_6$)—fully fluorinated non-carbon compounds—as potential technosignature gases (Fig. 1). A fully fluorinated compound is a molecule where the central atoms are only bonded to fluorine atoms. For example, in $NF_3$ the central nitrogen atom can bind to three other atoms. Because each of the bound atoms is fluorine, we call $NF_3$ fully fluorinated. $NF_3$ and $SF_6$ have been only briefly mentioned as technosignatures[34,35], the case has not previously been developed.

We are motivated to explore non-carbon fully fluorinated compounds primarily because life on Earth never makes compounds containing the N–F and S–F bonds, nor fully fluorinated non-carbon compounds ("Results" section). S–F and N–F bonds are not even made as metabolic intermediates and are likely to be universally excluded by life, no matter its biochemistry. In fact, life on Earth very rarely uses F chemistry. Only a few species make C–F bonds at all (Petkowski et al. in prep.). While life on Earth does not make fully fluorinated carbon molecules (e.g., tetrafluoromethane, $CF_4$, or CFCs that have been previously considered as potential technosignature gases[12]), life could in principle do so without inventing completely new enzymes and other necessary biochemical repertoire (Petkowski et al. in prep.). Hence fully fluorinated non-carbon molecules may be more robust technosignature gases than (fully) fluorinated carbon molecules.

$SF_6$ and $NF_3$ on Earth are not only industrial pollutants but their relative atmospheric abundance has rapidly increased[36,37] (See Fig. 2 and SI). $NF_3$ has no known abiotic sources and was entirely absent from the pre-industrial atmosphere[37]. Within the last half century, the $NF_3$ abundance in the atmosphere rose to close to 3 part-per-trillion (ppt) by volume (Fig. 2). $SF_6$ does not have significant abiotic sources that could mimic its rapid increase in the atmosphere. Like $NF_3$, the steady and rapid relative increase of $SF_6$ in the atmosphere from very low background abundances of <0.06 ppt to around 11 ppt in the last half a century combined with its relatively long atmospheric lifetime of at least a couple of hundred years[38] further supports $SF_6$ as a good technosignature gas candidate. However, we caution that the atmospheric chemistry of $NF_3$ and $SF_6$ has not been well studied and many potential destruction pathways for $NF_3$ and $SF_6$ may not be known (see Tables S2 and S3).

## Results

**No fully fluorinated molecules are made by life.**   Life on Earth is not known to produce any molecules with N–F or S–F bonds, and this includes fully fluorinated N and S compounds. We derived this result from our natural products database which is a curation of all known biochemicals and natural products (i.e. compounds produced by life) from an extensive literature online chemical repository search ("Methods" section and[39,40]). Here, "natural products" means chemicals made by life.

Life does produce some compounds with N–Cl, N–Br, S–Cl, and S–Br bonds, but none are volatile. In addition, the N–Cl, N–Br, S–Cl, and S–Br compounds are typically intermediaries and not molecules that accumulate on their own. The molecules containing N–Br and N–Cl are quite reactive and therefore rare, a notable example being a neurotransmitter N-bromotaurine[41] and pseudoceratonic acid[42]. The S–Cl and S–Br bond-containing molecules have been found only in proteins as intermediaries in synthesizing the N–S bonds[39].

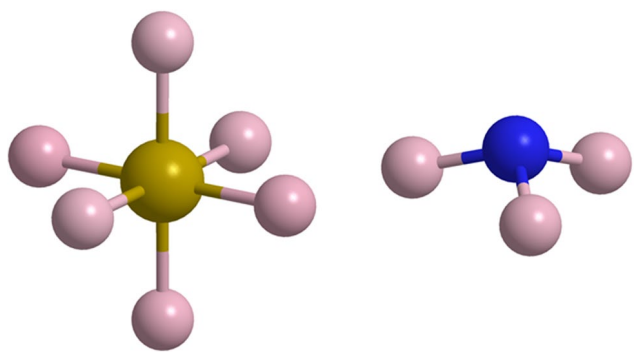

**Figure 1.** Chemical schematic of the fully fluorinated molecules: $SF_6$ (left) and $NF_3$ (right). Atoms are depicted by colors as follows. Yellow: sulfur. Blue: nitrogen. Pink: fluorine.

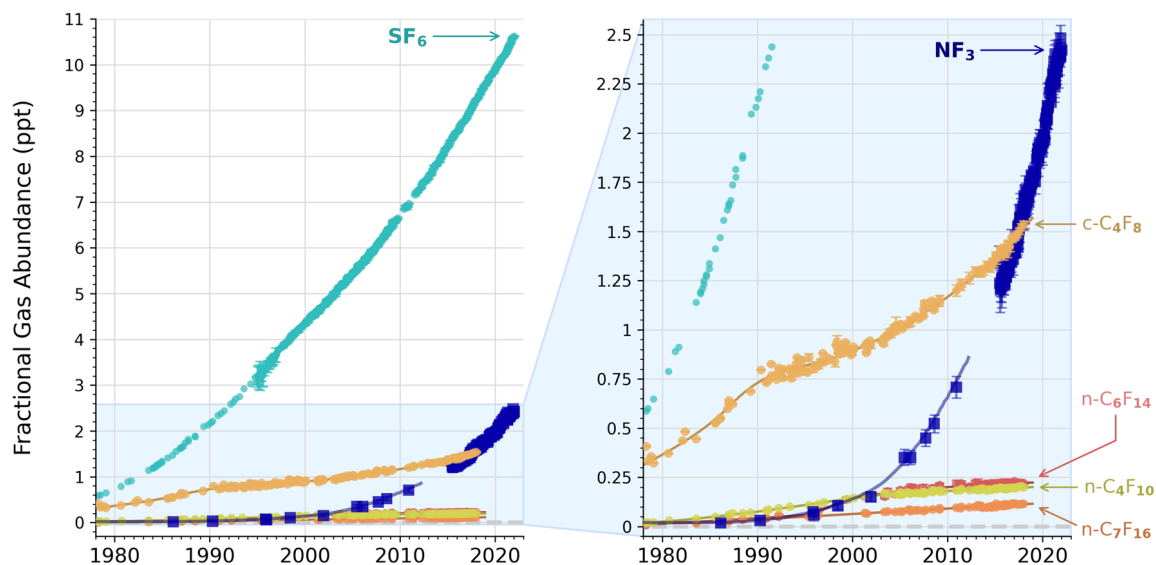

**Figure 2.** Atmosphere gas abundance for some industrial pollutants including $NF_3$ and $SF_6$. The y-axis is the fractional gas abundance in part-per-trillion (ppt) by volume and the x axis is time in years. Both $NF_3$ and $SF_6$ have a rapid increase compared to other industrial pollutants. The right panel is a zoomed in version of the left panel. Data from[36,37,95], and notably 2013–2022 data is from the Global Monitoring Laboratory (GML) ($SF_6$: https://gml.noaa.gov/hats/combined/SF6.html; $NF_3$: https://gml.noaa.gov/hats/gases/NF3.html).

In comparison with the rare N–Cl, N–Br, S–Cl, and S–Br containing molecules, there are thousands of compounds containing C–Cl, C–Br and C–I bonds that are made by life (Fig. 3). Most of the life-produced halogenated compounds are Cl-containing compounds (~ 2% of all known natural products, where the current known total number of unique natural products is ~ 220,000). Br-containing compounds produced by life are a close second (~ 1.7% of all known natural products). Iodine-containing natural products are much more rare but still a significant group with approximately 200 known examples (~ 0.1% of all known natural products). The above includes solids, liquids, and gases; it is worth noting that all of the volatile halogenated (Cl-, Br, I-containing) compounds produced by life are halocarbons, where the halogen atom is directly bonded to carbon.

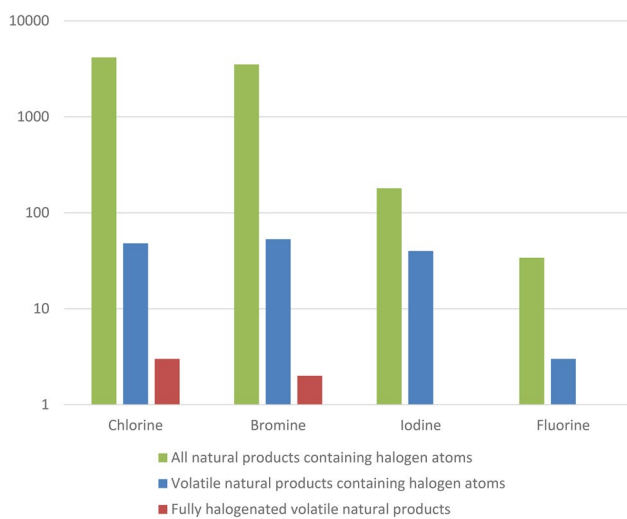

**Figure 3.** Number of molecules containing C–X bonds produced by life (called "natural products"), where X is Cl, Br, I, or F. For comparison the numbers are separated into three categories. Green: all known natural products in the category (i.e. produced by life). Blue: the subset of volatile natural products, here limited to molecules with 6 non-H atoms or less. Red: the subset of fully halogenated volatile natural products, also limited to molecules with 6 non-H atoms or less. Life rarely produces fully non-F halogenated volatile natural products and does not produce any fully fluorinated products of any kind (C–F, N–F, S–F, or other). Note that one fully halogenated molecule is double counted as it contains both Cl and Br, bromotrichloromethane. Not shown is that life does not produce any molecules with N–F or S–F bonds.

In contrast to the thousands of Cl-, Br- and I-containing carbon compounds made by life, the F-containing compounds are nearly excluded from life's repertoire, numbering only 34. And, only two of these 34 are volatile. Nearly all of the known biogenic F-containing natural products are fluorinated carboxylic acids (Petkowski et al. 2023 in prep.). There are no fully fluorinated C–F compounds known to be produced by life.

Although life does not produce fully fluorinated molecules, life does actually produce at least four carbon compounds that are fully halogenated with halogens other than fluorine. This is a very small percentage of all volatile halocarbons produced by life. For some numbers, and considering molecules with 6 or fewer non-hydrogen atoms, there are 85 volatile halocarbons produced by life. There are 34 possible halomethanes, of which 14 are fully halogenated. 22 halomethanes are produced by life on Earth[40]. Out of those 22, only 3 are fully halogenated: tetrachloromethane ($CCl_4$) produced by several plants and marine algae[43], tetrabromomethane, $CBr_4$, produced by from various marine algae, e.g. *Asparagopsis taxiformis*[44], and bromotrichloromethane, $CBrCl_3$, that contains both Cl and Br atoms, produced by marine algae[44]. For completeness, the fourth fully halogenated carbon compound produced by life is tetrachloroethene ($C_2Cl_4$), produced by Hawaiian red seaweed *Asparagopsis taxiformis*[44].

Again, the number of fully fluorinated molecules made by life is zero, no matter if they are fluorocarbons or if the F atom is attached to a non-carbon element.

We explain why life avoids F-containing compounds in Petkowski et al. 2023 (in prep.) and briefly summarize the explanation here. We have identified three challenges that F chemistry poses for life on Earth that make the use of fluorine in Earth's biochemistry a difficult prospect:

1. Relatively low bioavailability of F, which is primarily locked inside insoluble minerals, and is not available in surface water (unlike Cl, Br and I).
2. The uniquely high electronegativity of F, which means that enzymes that handle Cl, Br, peroxide and other oxidizing species cannot be repurposed to handle F. As a result of this challenge life needs to evolve a completely novel enzymatic machinery to create C–F bonds.
3. The lack of reactivity of the C–F bond, which makes evolving catalysts that can handle C–F bonds a difficult task.

These substantial evolutionary barriers mean that almost all life has found ways to address its ecological requirements with chemistry other than C–F chemistry. All three factors are general properties of fluorine chemistry, are not specific to terrestrial biochemistry and therefore are likely universal.

### $NF_3$ and $SF_6$ have unique spectral features compared to dominant atmospheric gases.

The gases $NF_3$ and $SF_6$ have unique spectral features as compared to bulk terrestrial planet atmosphere gases (Figs. 4 and 5). $NF_3$ and $SF_6$ absorption fall in the 9–12 micron spectral window where the plausible dominant super Earth or Earth-sized planet atmospheric gases $CO_2$, CO, $CH_4$, and the strong $H_2O$ vapor spectral features do not appear. Recall the gases $N_2$ and $H_2$ have no distinctive spectral features at infrared wavelengths; being homonuclear they have no net dipole moment. We note that the $NF_3$ and $SF_6$ strong absorbing power in a unique part of the spectrum compared to other major atmospheric gases is why they are potent greenhouse gases here on Earth. Trace gases of interest (such as $PH_3$, $NH_3$, $SO_2$, $H_2S$) other than the dominant terrestrial planet gases also do not have overlapping spectral features to $NF_3$ and $SF_6$ (Fig. 4).

There are gases with overlapping spectral features to $NF_3$ and $SF_6$, and these are primarily halogenated carbon compounds (Fig. 4). The multitude of possible halogenated carbon compound gases (which could be biosignatures or technosignatures) pose a more complicated situation. Disentangling $SF_6$ and $NF_3$ from the multitude of possible chlorofluorocarbon gases and those gases from each other, depends on spectral resolution. For $SF_6$ there are only a few candidates with similar main spectral peaks at the same wavelengths to $NF_3$ and $SF_6$. More work is needed to ascertain what is needed to distinguish amongst all of the halogenated gases, considering spectral features of other molecules.

### Atmospheric concentrations needed for detection.

On the order of 1 part-per-million (ppm) atmospheric abundance by volume of $NF_3$ and $SF_6$ produces a ~ 30 ppm signal in simulated transmission spectra for a terrestrial planet transiting an M dwarf star (Fig. 5). This is for an $H_2$-dominated atmosphere. For a $CO_2$- or $N_2$-dominated atmosphere, the signal is much lower, owing to the much smaller scale height for high mean molecular weight gases compared to the low mean molecular weight gas $H_2$ (Fig. 5). This finding is consistent with other biosignature gas detection simulations, including the point that detecting a ~ 30 ppm signal will almost certainly take tens of transits or more (e.g.,[2,45–48]), a much larger number than the typically few transits currently allocated for exoplanets with the JWST[29]. For details of simulated detectability including JWST noise floor see, for example[4,49,50].

However, we again emphasize a major point in favor of $NF_3$ and $SF_6$ spectral distinguishability in an exoplanet atmosphere is that the 9–12 micron region has no expected major atmosphere gases with strong spectral features—although many trace gases, especially halogenated compounds, have features in this window.

The atmospheric accumulation of $NF_3$ and $SF_6$ is favorable due to their low water-solubility (for a comparison to other gases such as $CO_2$ and $NH_3$ see Fig. 6). The low water solubility means they will not dissolve in rainwater and fall to the ground or the sea. The $NF_3$ and $SF_6$ and long atmospheric lifetimes also favor their accumulation (for destruction rates see the SI).

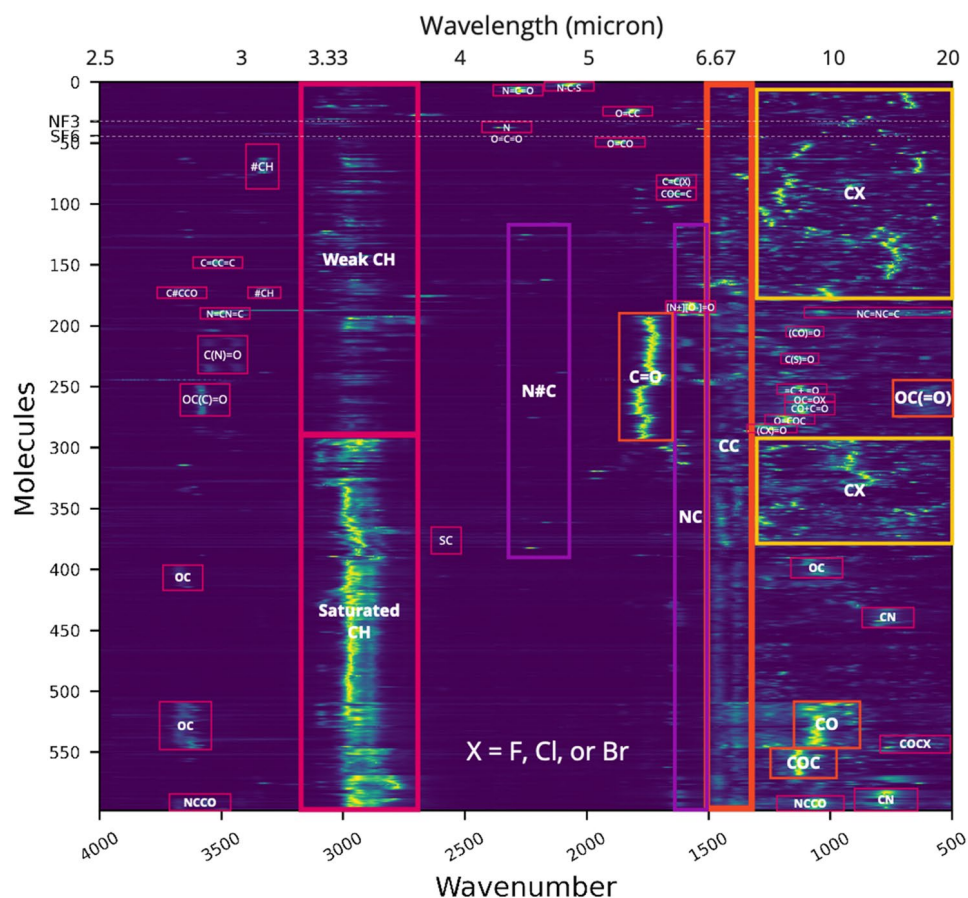

**Figure 4.** The molecular spectra phalanx plot compares the absorbance spectra amongst molecules. The x axis is wavenumber ranging from 4000 to 500 cm$^{-1}$ (2.5–20 μm). The y axis is the order of the molecules. Color represents the intensity of the absorption peaks; yellow and green represent strong absorption, while blue and purple represent weak or no absorption. Note the absorbance is normalized to 1. The spectra phalanx provides a visualization of which molecules are clustered together in wavelength based on their spectral data. Functional groups are labeled at their clustering points. The grey dashed lines near the top of the plot mark $NF_3$ and $SF_6$ and show their relative uniqueness in wavenumber space as compared to other halogenated species. The spectra data are from 600 gas species in our All Small Molecules Database (ASM)[40] (volatile molecules with up to 6 non-H atoms) that have available spectra in NIST[64]. See "Methods" for details on construction of this figure.

Proper estimation of the lifetime of $NF_3$ and $SF_6$ would require experimentally measured kinetics of chemical reactions of $NF_3$ and $SF_6$ in $H_2$ at various temperatures. The current thermochemical literature and databases (such as NIST) have very scarce information on reactivity of those species with relevant atmospheric components, at relevant temperatures (Table S1 and S2). Measurement or detailed modeling of those values is essential for progress.

Nonetheless, we can make some estimates. Regarding the lifetime of $NF_3$ in an $H_2$-dominated atmosphere, the rate constant [cm$^3$ molecule$^{-1}$ s$^{-1}$] for the reaction $H + NF_3 \rightarrow NF_2 + HF$ at 300 K has been calculated to be around $2.4 \times 10^{-20}$[51], which is four orders of magnitude lower than the rate constant of the reaction with the hydroxyl radical (OH), $4.0 \times 10^{-16}$, (reaction $OH + NF_3 \rightarrow F + H_2O + NO_2$ in Table S1). This difference suggests that the reaction with OH would dominate over the reaction with an H radical as a main destruction pathway for $NF_3$. As a result, the likely lifetime of $NF_3$ in an $H_2$-dominated atmosphere would not be significantly different than it is in the oxidized, OH-rich, atmosphere of Earth. Both Earth's atmosphere and the reduced $H_2$-dominated atmosphere are expected to be abundant with OH radicals due to photolysis of $H_2O$.

Regarding the lifetime of $SF_6$, it is a stable and unreactive gas with a lifetime of hundreds to thousands of years in Earth's oxidizing atmosphere. $SF_6$ likely has a similarly long, if not longer lifetime in a $H_2$-dominated exoplanet atmosphere. Experimental shock tube dissociation studies of $SF_6$ in the presence of $H_2$ gas in the temperature range of 1734–1848 K supports this conclusion. The measurements show that $H_2$ does not increase the dissociation of $SF_6$ as compared to argon gas control[52]. This result indicates that $SF_6$ should have at least a similar lifetime in the $H_2$-dominated atmosphere as it has in Earth's atmosphere.

**Abiotic sources and a false positive mitigation strategy.** *$NF_3$ has no known abiotic sources.* $NF_3$ has no known abiotic sources. In other words, $NF_3$ is not known to be a product of any photochemical, volcanic, or other geological process. We further show formation of $NF_3$ is thermodynamically unfavorable for terrestrial planet conditions (Fig. 7). $NF_3$ is also not known to be released from any fluorine-containing minerals. The absence of false-positives for $NF_3$ is supported by the lack of detection of this gas in any available pre-industrial samples[37]. With no know abiotic or biotic sources, $NF_3$ is a, if not the, prime candidate for a technosignature gas search.

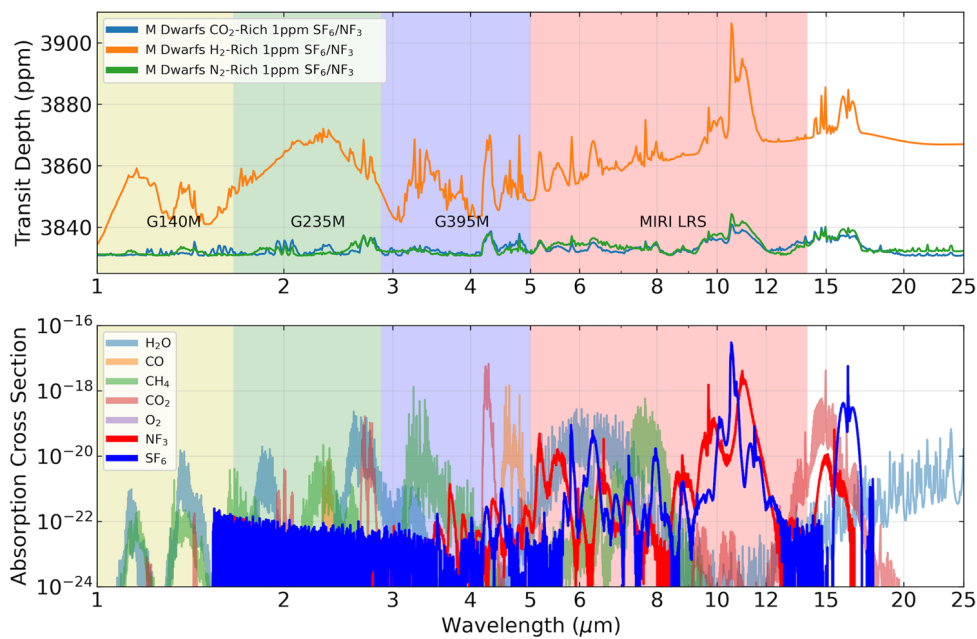

**Figure 5.** Simulated spectra of an exoplanet with transiting an M5V star with an atmosphere abundance of 1 part-per-million of $NF_3$ and $SF_6$. Top panel: the y-axis shows transit depth (ppm), and the x-axis shows wavelength (μm). The spectra are simulated from 1 to 23 μm, covering the wavelength span of most of JWST's observation modes. The yellow, green, and blue regions show the spectral coverage of NIRSpec, and the red region shows coverage of MIRI LRS. Bottom panel: absorption cross sections in cm$^2$ as a function of wavelength for key molecules of interest. For context, the contemporary Earth abundance levels are 3 ppt[37] and 11 ppt[38] for $NF_3$ and $SF_6$ respectively.

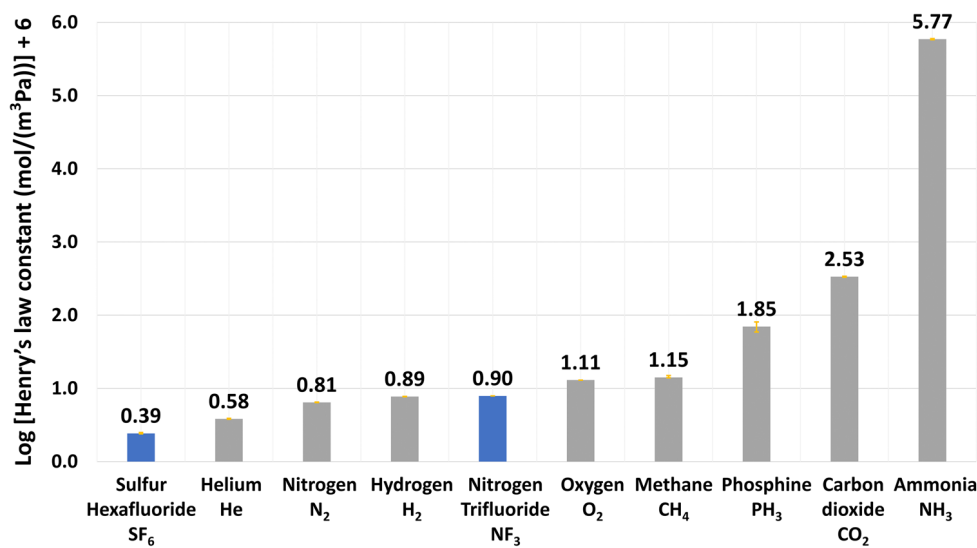

**Figure 6.** Solubility of various atmospheric gases in water. The x-axis shows the chemical species' names, and the y-axis shows the Henry's law constant on a log scale. Henry's law is defined as $H^{CP}_{(X)} = C(X)/p$, where $H^{CP}_{(X)}$ is Henry's law constant for a species X in mol Pa$^{-1}$ m$^{-3}$. P is the partial pressure of that species in Pascal, and C(x) is the dissolved concentration (in mol m$^{-3}$) under the equilibrium condition. The larger $H^{CP}$, the more soluble the species is. $NF_3$ and $SF_6$ have very low water solubility.

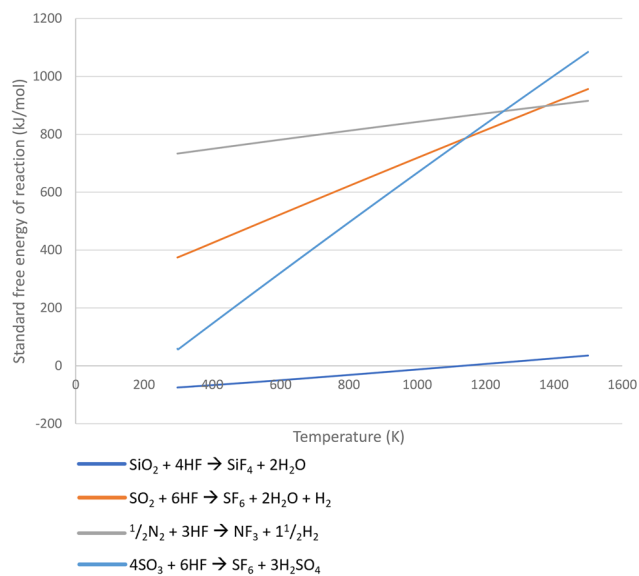

**Figure 7.** The free energy of formation of $NF_3$, $SF_6$, and $SiF_4$. The y-axis is the standard free energy of formation and the x-axis is temperature (K). The unfavorable thermodynamic conditions (positive ΔG) of the formation of $SF_6$ and $NF_3$ at all relevant temperatures are in contrast to the favorable abiotic formation of $SiF_4$ (negative ΔG), where only $SiF_4$ is a known volcanic product on Earth. The formation of $NF_3$ and $SF_6$ is highly thermodynamically unfavorable and therefore unlikely to be anything but a trace product of planetary geology or volcanism.

*Abiotic sources of $SF_6$.* On Earth trace amounts of $SF_6$ exists in volcanic rocks in rift zones, faults, igneous intrusions, geo-thermic areas and diagenetic fluids[53]. $SF_6$ is predominantly present in fluorites and some granites, while, for example basalts do not contain detectable $SF_6$[53,54].

The exact mechanism of abiotic formation of $SF_6$ on Earth is unknown. It is also unclear if $SF_6$ is directly made by volcanoes on Earth or its release is just associated with volcanic activity. Harnisch and Eisenhauer have examined the gases from several volcanic fumaroles, e.g., Etna (Sicily), Vulcano Island (Sicily), Kuju (Japan), and Satsuma Iwojima (Japan), and find that they are not significant sources of $SF_6$[54]. However, the authors note that the underlying rocks of these volcanoes are not granitic and as a result might lack a source for $SF_6$[54]. The equilibrium pre-industrial atmospheric concentration of $SF_6$ on Earth is estimated to be < 0.06 ppt[53]. The dominant F-containing volcanic gas on Earth is HF with abundances reaching 0.5–15 ppb[55]. Other trace F-containing species, including $NH_4F$, $SiF_4$, $(NH_4)2SiF_6$, $NaSiF_6$, $K_2SiF_6$, $KBF_4$, and organo-fluorides, are also associated with volcanic activity or released by volcanoes, but to a much lower extent than HF[56–59]. It is therefore unlikely that $SF_6$ will be a significant false-positive on a water-rich terrestrial planet as Earth.

The hypothesis that volcanic $SF_6$ is negligible on Earth is also supported by thermodynamics of $SF_6$'s formation (Fig. 7). The formation of $SF_6$ (and $NF_3$) is highly thermodynamically unfavorable and therefore unlikely to be a source of anything but a trace product of planetary geology or volcanism. The unfavorable thermodynamic conditions of the formation of $SF_6$ (and $NF_3$) are in contrast to the abiotic formation of $SiF_4$, another non-carbon fully fluorinated gas, that is a known volcanic product on Earth[60] (Fig. 7).

*Strategies to rule out $SF_6$ (and $NF_3$) false positives.* We first ask if $SF_6$ could be a significant abiotic gas on a planet with an environment different from Earth. The answer is yes, as follows. On a dry planet, or a planet that is otherwise H-depleted one would expect a different profile of F-containing volatiles erupted by volcanoes than on Earth which could lead to a potential false-positive interpretation of the source of the detected $SF_6$. On a H-depleted planet, fluorine will be bound to a greater extent to elements other than H. Therefore HF, while still expected to be erupted by volcanoes, would not dominate volcanic gases. Such a scenario opens the possibility for $SF_6$ to be a much more abundant volcanic product on dry exoplanets than it is on wet planet Earth.

The view that $SF_6$ could be a volcanic product on an H-depleted world is supported by the tentative detection of 0.2 ± 0.1 ppm $SF_6$ in the atmosphere of Venus by Venera 14[61]. If correct, this value is five orders of magnitude larger than the amount of $SF_6$ detected in Earth's atmosphere. Since Venus' crust and atmosphere are significantly H-depleted (though the deeper mantle may be relatively less H-depleted[62]), it is likely that the majority of F is erupted as other compounds than HF, e.g. $SSF_2$, $COF_2$, $FClCO$, and $SOF_2$, etc.[63]. Therefore, it is not unexpected that in the H-depleted environment of Venus, with abundant sulfur, $SF_6$ could also be a volcanic product released in significantly higher abundance than on Earth. $SF_6$ could also be the result of weathering of fluorite minerals which abundance on Venus is poorly constrained.

We now turn back to non-H-depleted planets that are the focus of this paper, and propose a strategy to rule out any volcanic origin of $SF_6$. We propose simultaneous observations of $SiF_4$ to be employed as a method to rule out volcanic sources of $SF_6$ (and $NF_3$). The overview reason is that $SiF_4$ is much more thermodynamically favored over $SF_6$ (see SI Section 4) such that any volcanic activity that produces $SF_6$ will produce significantly higher amounts of $SiF_4$. In more detail, $SiF_4$ is a known volcanic gas on Earth. Volcanic production of $SiF_4$ can reach several tons per day and in some instances, such as in the Satsuma-Iwojima volcano plume, $SiF_4$ production can rival that of HF[60]. As with $SF_6$, the formation of $SiF_4$ will be favored in hydrogen-depleted environments and low temperature gas sources. It is conceivable that on a planet with much more active low-temperature volcanism and lower crustal and atmospheric H abundance than Earth that a larger quantity of $SiF_4$ would be released by volcanoes into the atmosphere. $SiF_4$, therefore, can be an indicator of geological activity on a planet with volcanic chemistry much more favorable to forming $SF_6$ than Earth.

In an event where $SF_6$ is detected but simultaneous observations of $SiF_4$ is not, the likelihood increases that $SF_6$ is biological or technological. This conclusion is supported by our calculations that suggest that there are no plausible conditions (pressure 0.1–10,000 bar, $H_2O$ content 0.1–95%, mantle redox state (MH vs. IW etc.)) where the amount of $SF_6$ (or $NF_3$) produced by volcanoes comes to within ten orders of magnitude of that of $SiF_4$, effectively ruling out the possibility of volcanically-driven co-existence of $SiF_4$ and $SF_6$ (or $NF_3$) (Figure S1 and Figure S2). All of the above rationale applies to $NF_3$; unlike $SF_6$ there are no abiotic sources on Earth, and so volcanism is even less likely to make this gas on another planet.

There appears to be no spectral information spectral information for $SiF_4$[64], getting such information is crucial for the execution of our proposed mitigation strategy. We conclude this section with a call to study the spectroscopy of the fully fluorinated non-carbon molecules.

## Discussion and summary
**Prospects.** The JWST has been operational for science as of 2022 and is our best currently existing capability for exoplanet atmosphere observations for transiting planets via transit transmission spectroscopy. Two categories of terrestrial planet atmospheres are accessible by JWST. The first is the not yet existing terrestrial planets transiting white dwarf stars (See "Introduction" section). The second category is terrestrial planets transiting small red dwarf stars. In addition to the long-required observation times for technosignature gases for such systems ("Introduction" section), a major challenge is the red dwarf host star variability (e.g.[65] and references therein). Stellar variability is a collective term for a rich set of physical phenomena caused by stellar surface inhomogeneities which come in the form of granulation and magnetic features such as spots and faculae (e.g.,[66]). The host star variability changes with time as active regions evolve and as the star rotates. Stellar spots and faculae have different temperatures from the disk-averaged photosphere, and for cooler stars, have molecular features distinct from the star itself but similar to those in a planet's atmosphere ([65] and references therein). Stellar surface inhomogeneities can induce rogue features in stellar data which mimic signatures of exoplanet atmospheres; some studies have found that the signal from stellar inhomogeneities exceeds the signal from the planetary spectral features (e.g.,[65,67,68]). A community of over 100 experts has summarized the challenges and potential mitigation strategies of M dwarf host star variability effects on small transiting exoplanet atmospheres (NASA's Exoplanet Exploration Program Study Analysis Group 21 (SAG21)[65]).

New large ground-based telescopes now under construction are planned to be online within the next decade: the Extremely Large Telescope (EELT, 39 m aperture diameter)[69,70]; the Thirty Meter Telescope (TMT, 30 m aperture diameter)[71]; and the Giant Magellan Telescope (GMT, 20 m aperture diameter)[72,73]. These large telescopes can directly image habitable-zone planets orbiting (i.e., not necessarily transiting) M dwarf stars with the right coronagraph instrumentation and extreme adaptive optics. The challenge is overcoming the high planet-star contrast at $10^7$–$10^8$ levels. METIS[74] on the EELT will have mid-infrared direct imaging capability via a coronagraph and extreme adaptive optics that includes low and medium resolution spectroscopy; however METIS is designed for planet discovery via imaging and its limited sensitivity means METIS is unlikely to enable reliable ~ 10 micron spectrum of a temperate, rocky world orbiting an M dwarf star ([74] and Quanz, priv. comm. 2023). Direct imaging in the thermal infrared for about five Sun-like stars may also be possible[74,75]. Near infrared imaging and spectroscopy of reflected light from rocky planets in their host star's habitable zone is anticipated to be possible for up to 100 nearby low-mass stars (primarily mid M dwarf stars)[75-77]; while the near-IR is outside of the wavelength range of interest for $NF_3$ and $SF_6$, such programs may help discover new suitable planets for follow up atmosphere observations. Other than direct imaging, the large ground-based telescopes might be capable of a combination of high-dispersion, high-spectral resolution (R ~ 100,000) spectroscopy with moderate high-contrast imaging to observe spectra of a few rocky planets orbiting Sun-like stars[78]. (Note that not all IR wavelength regions are not easily accessible from Earth's surface due to Earth's atmospheric gases).

NASA's Habitable Worlds Observatory (HWO; https://www.greatobservatories.org/irouv) is a NASA Great Observatory intended to be ready for launch in the mid-2030s and will be designed to directly image exoplanets orbiting Sun-like stars for both discovery and atmospheric characterization. This telescope will have a primary mirror about 6-m in diameter and is planned to have a coronagraph to block out starlight so the planet can be directly imaged. Because HWO will operate at visible to near-IR wavelengths, HWO will not be able to access the infrared windows (> 4 microns) appropriate for $NF_3$ and $SF_6$ spectral features.

The space-based interferometer under study called the Large Interferometer for Exoplanets (LIFE;[79]) is designed to observe at 4–18.5 microns, and this includes the infrared windows for $NF_3$ and $SF_6$ spectral features (Fig. 5). LIFE would have four elements each with aperture 2–3.5-m diameter as well as a combiner spacecraft[80].

**Are $SF_6$ and $NF_3$ alien biosignatures or universal technosignature gases?** The exclusion of N–F and S–F compounds from Earth biochemistry, the low water solubility of $NF_3$ and $SF_6$, their unique spectra, long atmospheric life time, and industrial utility on Earth all makes $NF_3$ and $SF_6$ attractive targets as technosignature gases.

We discuss the reasons for and against $NF_3$ and $SF_6$ being produced by an alien biology versus by a technological society.

Life on Earth has apparently never made an N–F or an S–F bond-containing molecule, despite life independently evolving F-handling enzymes several times over the last three billion years (Petkowski et al. 2023, in prep.). Life also rarely makes fully halogenated molecules, with only four known examples ("No fully fluorinated molecules are made by life" section). This suggests that there is at least one major evolutionary barrier to making $NF_3$ or $SF_6$, which would require a correspondingly powerful selective benefit to overcome.

$SF_6$ is a completely chemically inert gas, making it essentially "invisible" to biochemistry. Therefore, we have to ask why life, no matter its biochemical makeup, would invest significant metabolic machinery and energy into making a compound that it then throws away, and which, due to its chemical inertness, has a negligible effect on the organism's immediate environment or on its competitors. Biological production of $SF_6$ would cost an organism a large amount of energy while not giving any evolutionary advantage (i.e., does not provide any obvious fitness gain).

$NF_3$ is mildly reactive, and so we may speculate that $NF_3$ could have the same ecological role as $CH_3Br$ on Earth, as a mildly reactive, non-specific, diffusible toxin to intoxicate the competition in the immediate environment. Therefore, $NF_3$ production could increase the fitness of the $NF_3$-producing organism. The toxicity of $NF_3$ can be to an extent similar to the toxicity of phosphine ($PH_3$)[81]. $PH_3$ does not react all that much with biological tissues, but it is toxic to humans, and many other $O_2$-dependent organisms, because it converts hemoglobin to methemoglobin (which cannot bind oxygen), i.e. $PH_3$ is toxic to (large, terrestrial) oxygen-dependent organisms but not to anaerobic ones[82]. Oddly, in identifying a technosignature gas, a biological source may be a false positive. Against this, we note that there are many mildly reactive, toxic carbon-containing compounds that can be made with the enzymatic machinery that any carbon-based life is likely to possess, such as $CH_3Br$, $CH_3I$, cyanogen, CO, formaldehyde, nitric oxide, so $NF_3$ would not provide any unique advantage over these more evolutionarily accessible substances.

One could argue that the biological repertoire of gases on other planets may surprise us because we have no idea what gases non-Earth-like life might produce[40]. Production of certain biochemicals is often an evolutionary accident or depends on the planetary environmental history. An example is the gas stibine, $SbH_3$, which would not be expected to be made by Earth life because Sb itself is a rare element in the Earth's crust, but nevertheless is synthesized by terrestrial anaerobic sewage sludge microflora[83]. Another example is trimethylbismuth ($C_3H_9Bi$) produced by a variety of bacteria in anaerobic conditions (reviewed in[84]). But we favor the point that while it is possible that an alien biochemistry would find use for $NF_3$ and $SF_6$, virtually any combination of physical and chemical properties of $NF_3$ or $SF_6$ can be duplicated with less energy and less dangerous radicals using the chemistry of other halogens and carbon. Indeed, our premise of promoting $NF_3$ and $SF_6$ as technosignatures is that F is nearly excluded by life on Earth (Petkowski et al. 2023, in prep.)) and that such an exclusion may well be universal.

We further argue that industrial use of some chemicals may be likely to be universal. $SF_6$ in particular has unique properties that make it useful for a technological civilization (reviewed "Results" section) and specifically its high dielectric constant and high breakdown voltage as a gas make it valuable in high voltage electrical equipment[36]. Biosignature gases are the product of evolutionary contingency and are limited by thermodynamics and the reactivity of materials to water. Industrial chemicals, however, are the result of an informed, systematic search of all possible chemicals for materials that have optimal properties for a specific application, largely regardless of the thermodynamics of their synthesis or whether their synthesis requires chemistry that is incompatible with earth surface conditions (for example, industrial production of $NF_3$ is done in molten ammonium fluoride[85], an obscure material unlikely to occur on its own on any rocky planet). It is therefore plausible to suggest that an extraterrestrial civilization that wanted, for example, a gaseous product to act as a high voltage insulator and arc quencher, would choose $SF_6$, no matter what the entity's own biochemistry, evolutionary history or planetary environment was.

If we in the future have a way to detect the tiny part-per-trillion amounts elsewhere that we humans have accumulated in our atmosphere, and are lucky to catch the small window where a society becomes industrial, we may observe the rapid, steady increase of the atmospheric abundance of $SF_6$ or $NF_3$. A rapid, steady increase would favor a technological source and may be a solid discriminator between bio- and a true technosignature gas. Even the relatively rapid increase in atmospheric $O_2$ that led to the Great Oxygenation Event still took 1–10 million years for $O_2$ to accumulate in high enough concentrations to have a weathering effect on rocks (e.g.,[86]). A multi-generational observational campaign can further distinguish between $SF_6$ and/or $NF_3$ as technosignature and not biosignature gases. Observation of a planet over the time span of 2–4 generations (~ 100 years) to monitor the increase of the $SF_6$ and $NF_3$ gases could strengthen attribution to a technological source, but one would have to get lucky with timing.

We now turn to the requirement that the atmospheric abundance of $NF_3$ and $SF_6$ needed for detection with the JWST is of the order of 1 ppm, far higher than current atmospheric levels. Here on Earth, no uniquely technological gas has been produced to accumulate to such relatively high amounts in the atmosphere. 1 ppm $SF_6$ is 100 times the current terrestrial level, and it would take ~ 700 years of the current emission rate to build that concentration in Earth's atmosphere, taking the $SF_6$ atmospheric lifetime of 3200 years. The greenhouse gas effect of such concentrations on Earth would be catastrophic, which suggests that any alien technological society would curtail their emissions before they reached 1 ppm, unless their goal was substantial global greenhouse heating. Indeed $SF_6$ has been considered as a terraforming agent on Mars, albeit for humans at our stage of development a prohibitively expensive one[87].

Fully fluorinated non-carbon containing gases other than $NF_3$ and $SF_6$ have little known about them (see discussion in the SI; Table S3).

**Summary.** In summary, $NF_3$ and $SF_6$ are appealing technosignature gases primarily because use of F is nearly excluded by the chemistry of life on Earth for fundamental reasons (Petkowski et al. 2023, in prep.), and moreover because molecules with N–F and S–F bonds are not made by life on Earth at all. We have argued that while an alien biochemistry might find use for $NF_3$ and $SF_6$, such use would have to provide a significant evolutionary gain that offsets e.g. the large energy expenditure for the synthesis and breakage of fluorine-containing bonds. In contrast, industrial use may be universal due to unique physical and chemical properties of $NF_3$ and $SF_6$ gases (see SI). Therefore, $NF_3$ and $SF_6$ may be likely to be used by alien industry no matter the biochemical makeup of the alien biology or the particular environmental conditions of the alien planet. We note however that it is likely that the technological stage of the civilization, or time-span at which $NF_3$ and $SF_6$ are produced in sufficient amounts to be observed is short.

$NF_3$ has no known sources other than industrial, while $SF_6$ is produced abiotically in extremely small, trace amounts. Their source can be distinguished from e.g. potential volcanic release into the atmosphere by comparison with the known volcanic gas $SiF_4$. The lack of known false-positives for $NF_3$ and very low abundance of abiotic sources of $SF_6$ on Earth supports their potential as a technosignature gases on exoplanets. $NF_3$ and $SF_6$'s long atmospheric lifetimes and unique spectral features aid their potential detection.

## Methods

**Custom natural products database.** To understand the uniqueness of fully fluorinated compounds out of all chemicals made by life (called "natural products"), we use our database of natural product chemicals curated over the last decade. Our database is presented in[40] and expanded and completed as described in[39]. We created and curated our database by an extensive literature search and by searching available online natural product repositories[39].

Our natural products database has been rigorously screened to contain only compounds that are a result of natural biochemical processes of a living organism. It also contains biological sources identified for every molecule (i.e., a list of species from which the natural product was isolated).

We emphasize how challenging it is to compile a complete list of all that is known about each natural product. First, no individual database covers more than 20% of the known natural products. Second, because most natural product databases focus on drug design, they include synthetic derivatives of natural products or drug metabolites that often mimic natural molecules while not being true natural compounds themselves. Other databases often include artificial compounds that emerge as a result of "feeding" an organism with a precursor molecule, or completely artificial compounds that have accumulated as contaminants in plants and animals. We manually excluded such compounds from our database. Third, most data sources needed extensive checking and modification due to a range of format differences and coding errors. All of the above problems motivated us to curate our own natural products database (see[39] for a full description).

**Spectra visualization tool.** A key question for any atmospheric trace gas is whether or not its spectral features are distinguishable from features found in other expected atmospheric molecular gases. We introduce a new tool for analysis, detailed in[88] and summarized here.

We call our new tool a *spectra phalanx plot* (Fig. 4). This plot enables a visual comparison amongst the absorption peaks of each individual molecule's spectral features, with molecules that share similar spectral features grouped closer together. Each molecule occupies a horizontal line parallel to the x-axis, where the molecule's spectral features are plotted as a function of wavelength location. We use color to represent the intensity of the absorption peaks; yellow and green represent strong peaks, while blue and purple represent weak or no absorption. Note the absorbance is normalized to 1, so only the ratio between the peaks and the strongest peak is important, and comparing the absolute intensity between molecules is not possible. For this tool, we use the ~ 600 transmission/absorbance data spectra available from NIST[64] (with a small subset from HITRAN[89]). The order of the molecules is not meaningful other than that molecules with similar wavelengths of spectral features are grouped together.

We generate the molecule ordering in the spectra phalanx plot using hierarchical clustering, a tree-based approach that builds clusters by iteratively grouping two of the closest cluster/elements into the same cluster and organized in a binary tree structure called a dendrogram[90]. We apply hierarchical clustering on molecular spectra to cluster the molecules and validate that the molecules in the same clusters share similar chemical structures using a molecular maximum common substructure search using methods described in[88].

In more detail, comparing the molecular IR spectra of two molecules expresses how similar the two are. We use the squared Euclidean distance function to compare normalized molecular spectral signatures, each normalized to have unit peak absorbance. An agglomerative clustering process organizes the pairwise symmetric distance matrix between all pairs of molecules into a cluster hierarchy. Using the hierarchy, one can visualize and identify the molecular spectral features that contribute to the relative detectability of molecules. It is these spectral features that one would want to detect. To locate the distinguishing features, we enumerate the molecules row-by-row in the order of their cluster similarities, imaging a molecule's spectral signature. The one-dimensional image along the x-axis indicates a bright (yellow) color for high absorbance and dark (blue) for low absorbance (in contrast to a curve plot). Most interesting is the pattern that emerges when one looks at all cluster-enumerated molecular spectral signatures. This phalanx plot shown in Fig. 4 shows that normalized peaks (with some spectral width) at specific wavelengths are shared by many molecules, and these do not contribute significantly to the distinguishability and, thus, relative detectability of the molecule. But it also appears, as shown, that specific clusters share peak distributions that are relatively muted or absent in all the others. These are the distinguishing IR signature features that help detect the molecules in the group. To the degree that the number of molecules in a group is small and the spectral features shared by them are exclusive relative to the others in our All Small Molecule (ASM) database[40], they are highly detectable. See[88] for more details.

**Atmospheric spectra simulator.** We simulate model atmospheres to assess the approximate atmospheric abundance of gases needed for detection with simulated James Webb Space Telescope (JWST) observations. We use the computer model "Simulated Exoplanet Atmosphere Spectra" (SEAS) code from[88].

We use the molecular mixing ratio profiles and calculate the optical depth of each layer of the atmosphere[88,91]. We calculate the stellar intensity absorption along each path through the planet atmosphere by $A = n_{i,j}\ \sigma_{i,j}\ l_i$, where $A$ is absorption, $n$ is number density, $\sigma$ is the absorption cross-section and $l$ is pathlength. The subscript $i$ denotes each layer that the stellar radiation beam penetrates, and $j$ denotes each molecule. For the height of each atmospheric layer we adopt scale height of the atmosphere. We calculate the transmittance, $T$, of each beam using the Beer–Lambert Law. Then, we compute the total effective height $h$ of the atmosphere by multiplying the absorption $A = 1 - T$ by the atmosphere's scale height. To connect to observations, we calculate the total attenuated flux as transit depth $(R_{planet} + h)^2/R_{star}^2$ in units of ppm.

We simulate transmission spectra for a hypothetical 1.5 $R_{Earth}$, 5 $M_{Earth}$ super-Earth transiting an M dwarf-star similar to GJ 876. We choose 1.5 $R_{Earth}$ as it is consistent with a rocky planet[92]. We simulate three planetary atmospheres: ones dominated by $H_2$, $N_2$ and $CO_2$. We choose a relatively massive super Earth planet because a more massive planet is more likely to retain an $H_2$-dominated atmosphere than a lower mass planet. We choose to simulate a rocky exoplanet with an $H_2$-dominated atmosphere because such an atmosphere is favorable for detection by transmission spectroscopy than a higher mean molecular weight atmosphere such as one dominated by $N_2$ or $CO_2$. We also simulate an $N_2$ and a $CO_2$-dominated atmosphere. We take the temperature, pressure, and vertical gas abundance profiles from[48], where the molecular gas abundances were computed from a photochemical equilibrium model[93,94]. We assume varying atmospheric abundances of $SF_6$ and $NF_3$, from 1 ppb to 1 ppm.

## Data availability
The datasets used and/or analyzed during the current study are available from the corresponding author on reasonable request.

### Acknowledgements
We thank Lauren Herrington for creating Figure 2.

### Author contributions
Conceptualization: S.S.; Methodology S.S., Z.Z., J.H., W.B, S.R., J.J.P.; Formal analysis: S.S., Z.Z., S.R., J.H., W.B, J.J.P.; Writing—original draft preparation: S.S. and J.J.P.; Writing—review and editing: S.S., Z.Z., J.H., W.B, S.R., J.J.P.

### Funding
This work was in part supported by the Heising-Simons Foundation Grant 2018-1104 and by NASA Grants 80NSSC19K0471 and 80NSSC20K0586.

### Competing interests
The authors declare no competing interests.

### Additional information
**Supplementary Information** The online version contains supplementary material available at https://doi.org/10.1038/s41598-023-39972-z.

**Correspondence** and requests for materials should be addressed to S.S.

**Reprints and permissions information** is available at www.nature.com/reprints.

**Publisher's note** Springer Nature remains neutral with regard to jurisdictional claims in published maps and institutional affiliations.

Supplementary Information for

# Fully Fluorinated Non-Carbon Compounds NF$_3$ and SF$_6$ as Ideal Technosignature Gases

Sara Seager[1,2,3], Janusz J. Petkowski[1,4], Jingcheng Huang[1], Zhuchang Zhan[1], Sai Ravela[1], William Bains[1,5]

**Affiliations**

[1] Department of Earth, Atmospheric and Planetary Sciences, Massachusetts Institute of Technology, 77 Massachusetts Avenue, Cambridge, MA 02139, USA
[2] Department of Physics, Massachusetts Institute of Technology, 77 Massachusetts Avenue, Cambridge, MA 02139, USA
[3] Department of Aeronautics and Astronautics, Massachusetts Institute of Technology, 77 Massachusetts Avenue, Cambridge, MA 02139, USA
[4] JJ Scientific, 02-792 Warsaw, Poland
[5] School of Physics and Astronomy, Cardiff University, 4 The Parade, Cardiff CF24 3AA, UK

*Correspondence: seager@mit.edu

## SI 1 A Summary of Proposed Technosignature Gases

Multiple studies have examined the detectability of CFCs in exoplanet atmospheres (e.g.,[1,2]). For instance, [2] investigated the detectability of CFC-11 ($CCl_3F$) and CFC-12 ($CCl_2F_2$) on TRAPPIST-1e. Assuming a James Webb Space Telescope (JWST) Mid-Infrared Instrument low-resolution spectrometer (MIRI-LRS) noise floor of 10 ppm, they found that present-day Earth abundances of these two CFCs could be detected on TRAPPIST-1e, with about 100 hours of JWST observation time[2]. However, assuming a conservative JWST noise level of 50 ppm, they concluded that even CFCs five times the present-day Earth level would not be detectable by JWST, regardless of the observation time [2]. In another study, researchers investigated the detectability of CFC-11 ($CCl_3F$) on an Earth-sized planet transiting a white dwarf star similar to Beta Persei (commonly known as Algol)[1]. They found that CFC-11 at an abundance ten times the present Earth level could be detected by JWST with about 1.2 days of observation time[1]. While no Earth-sized planets transiting bright white dwarf stars have yet been detected, they remain promising candidates for atmosphere study if they exist and can be discovered.

Other proposed technosignature gases include the simultaneous detection of $NH_3$ and $N_2O$ in an atmosphere that also contains $H_2O$, $O_2$, and $CO_2$ as a signature of extraterrestrial agriculture[3]. The gas $NO_2$ as an atmospheric technosignature has been proposed as a sign of an industrial revolution, specifically combustion engines[4]. These technosignature gases, however, are far from unique as both the planet and life also produces them. Terrestrial agriculture is an exploitation of the biochemistry of Earth, and so its reliance of exogenous sources of $NH_3$ (as fertilizer) and its production of $N_2O$ might be specific to terrestrial biology, and not a general sign of agriculture. $NO_2$ from internal combustion engines assumes a very specific technological trajectory for the planet: the widespread use of a specific type of fossil-fuel-powered transport infrastructure rather than, for example, steam- or electric-powered transport.

## SI 2 An Overview of $NF_3$ and $SF_6$

Both $NF_3$ and $SF_6$ are present in Earth's atmosphere not from intentional release but from leakage from industrial use. Here we summarize the origin and properties of $NF_3$ and $SF_6$.

### 2.1 $NF_3$ and $SF_6$ Physical and Chemical Properties

At room temperature, $NF_3$ is a colorless and non-flammable gas. At room temperature, $NF_3$ is only slightly soluble in water[5] and does not react with water or dilute acids[6–8]. $NF_3$ is thermodynamically and chemically stable; for example it does not react with most metals below 250 °C[6,9], but it can act as a potent yet slow oxidizer[8].

In recent years, $NF_3$ has been widely used in the microelectronics and semiconductor industries where it is used as an etchant in producing thin-film-transistor liquid-crystal displays (TFT-LCD), semiconductors, and solar photovoltaic panels (e.g.,[6,9–12]).

Sulfur hexafluoride ($SF_6$) is a colorless, non-toxic, and non-flammable gas[7]. At room temperature, $SF_6$ is almost insoluble in water; the solubility of $SF_6$ is even lower than that of helium (He), one of the least water-soluble gases[5] (Figure 6 in the main text). $SF_6$ is completely chemically inert and non-toxic. At room temperature, it is unreactive towards most metals. It does not react with magnesium (Mg) and copper (Cu) even if they are heated to "red hot" temperatures. It does not react with phosphorus (P) or arsenic (As), nor with hydrochloric acid (HCl), sodium hydroxide (NaOH), or potassium hydroxide (KOH). $SF_6$ does not decompose even when heated to 500 °C. In addition, $SF_6$ does not react with liquid water or high-pressure steam. $SF_6$ will only react with boiling sodium (Na)[7,13–18]. $SF_6$'s only toxic effect on mammals, including humans, is when it is breathed in at greater than 80% concentration and so replaces oxygen in inspired air[19].

The main reason behind the chemical stability and unreactivity of $SF_6$ is a kinetic barrier to its hydrolysis[7]. $SF_6$ has a unique molecular structure among sulfur-containing compounds. $SF_6$ is an octahedrally-shaped molecule with six fluorine atoms symmetrically attached to the central sulfur atom (Figure 1 in the main text). As a result, the S atom is sterically shielded by the surrounding F atoms, making it inaccessible to other reacting molecules, e.g., water.

$SF_6$ is used in a wide variety of industries, including the electrical power industry, semiconductor manufacturing, and the production of aluminum and magnesium, to name a few[7,20–22]. In the electrical utility industry, $SF_6$ is often used as an insulating gas in electrical transmission and distribution equipment, such as current/voltage transformers, circuit breakers, switchgear, and capacitors (e.g.,[7,20,23]). $SF_6$'s chemical stability, non-flammability, and non-toxicity add to its usefulness as an excellent insulating gas[7,24].

As a result of the low reactivity of $NF_3$ and the chemical inertness of $SF_6$, once released to the atmosphere these gases have very long residence times.

**2.2 A Steady and Rapid Increase of Atmospheric Concentration of $SF_6$ and $NF_3$ in an Industrialized World**

The widespread use of $NF_3$ and $SF_6$ due to their unique advantages and pivotal roles across multiple industries has nonetheless raised concerns about their environmental impact. Both $NF_3$ and $SF_6$ are very potent greenhouse gases (e.g.,[25–27]). $NF_3$ has a 100-year Global Warming Potential (GWP) of about 16100, while $SF_6$ has a GWP of about 23500[28]. $NF_3$ and $SF_6$ are more than five orders of magnitude more efficient greenhouse gases than $CO_2$. Since $NF_3$ and $SF_6$ are very stable, they are difficult to break down and remove from the atmosphere, and hence $NF_3$ and $SF_6$ have very long atmospheric lifetimes. The atmospheric lifetime of $NF_3$ is about 500 years, and that of $SF_6$ is 850 - 3200 years[28,29].

Since the use of $NF_3$ in specialized industry began in the 1970s, and was later used in more wide-spread applications in the electronic industry in the late 1990's, the amount of atmospheric $NF_3$ has been rapidly increasing and has doubled every five years since the late 20th century (https://gml.noaa.gov/hats/gases/NF3.html). In contrast to $SF_6$, $NF_3$ does not have any known non-human-made source and measurements of air entrapped in ancient ice Antarctica as well as measurements of $NF_3$ in archived air tanks filled before 1975, i.e., before the introduction of $NF_3$ to human industry, have resulted in undetectable levels of $NF_3$. The undetectable levels of $NF_3$ in pre-industrial samples suggest that background $NF_3$ levels were essentially zero or no greater than the limit of detection of 0.008 ppt[30]. The rapid rise in the global atmospheric abundance and projected future demands of human industry place $NF_3$ as the fastest growing contributor to radiative forcing of all the synthetic greenhouse gases by the mid-XXI century[31]. Lack of detectable pre-industrial $NF_3$ supports its potential as a technosignature gas and strongly suggests that $NF_3$ does not have any significant abiotic source.

The increase in atmospheric $SF_6$ follows the same trend as $NF_3$. Over the last 70 years, since its first application in industry in 1953, $SF_6$'s concentration in the troposphere increased dramatically from approx. 0.05 ppt to > 4 ppt (Figure 2 in the main text)[32].

There is limited atmospheric chemistry reactivity data for $NF_3$ and $SF_6$ with potential destruction pathways possibly unknown. The known rates show very low reaction rates with the dominant atmospheric radical OH, and the only significant rates are for gases of low atmosphere abundance (Tables S1 and S2).

# SI 3 Other Fully Fluorinated Non-Carbon Molecules?

Given how compelling $NF_3$ and $SF_6$ are as technosignature gases, one should ask if other volatile fully fluorinated molecules are equally promising as technosignature gases.

Other fully fluorinated human-made gases do exist and are worth further exploration as technosignature gases because they do not exist in nature. However, there are a limited number of possibilities, and the ones that have been synthesized have limited data (e.g., reactivity, water solubility, or gas phase spectral feature information). The list includes $PF_3$, $P_2F_4$, $SeF_6$, or $S_2F_{10}$ (Table 1)[33]. Out of the list of possibilities (Table S3), $SiF_4$ is made by Earth's volcanoes and therefore not a suitable technosignature gas (see 5.3) and $PF_3$ slowly hydrolyzes in water. Despite its reactivity to water, $PF_3$ is worth further exploration, but so far, perhaps because it is not useful for industry, there is almost no spectral or any other relevant information on its potential as a technosignature gas. The other gas candidates listed in Table S3 are all highly reactive in water and not worth further comment, except to note that for some gases life does produce the hydrogenated versions, such as $NH_3$, $H_2S$, $PH_3$, $GeH_4$, but never partially fluorinated versions.

| Reactants | Products | Rate Law [cm³/molecule s] | Valid Temperature Range [K] | Rate at 298K | Source | Note |
|---|---|---|---|---|---|---|
| $NF_3 + N$ | $NF_2 + NF$ | $2.13 \times 10^{-12} (T/298)^{1.97} \exp(-15154.3/T)$ | 200 - 4000 | $1.98 \times 10^{-34}$ | NIST | |
| $NF_3$ | $NF_2 + F$ | $6.81 \times 10^{-8} \exp(-24174.7/T)$ | 1100 - 1800 | N/A | NIST | High temp rxn |
| $OH + NF_3$ | $F + H_2O + NO_2$ | $<4.0 \times 10^{-16}$ | 298 | $4.0 \times 10^{-16}$ | NIST | Products unknown |
| $CF_3 + NF_3$ | $CF_3NF_2 + F$ | $1.99 \times 10^{-14} \exp(-3740.5/T)$ | 303 - 423 | N/A | NIST | |
| $CH_3 + NF_3$ | $CH_3F_2N + F$ | $7.09 \times 10^{-14} \exp(-4899.9/T)$ | 374 - 467 | N/A | NIST | |
| $C_2H_5 + NF_3$ | $C_2H_5F_2N + F$ | $2.41 \times 10^{-11} \exp(-8300/T)$ | 410 - 486 | N/A | NIST | |
| $O(1D) + NF_3$ | $NF_2 + OF$ | $1.1 \times 10^{-11}$ | 298 | $1.1 \times 10^{-11}$ | NIST | |
| | $O(3P) + NF_3$ | $1.40 \times 10^{-12} \exp(44/T)$ | 199 - 356 | $2.3 \times 10^{-11}$ | JPL 19-5 | |
| | Products | $1.86 \times 10^{-11} \exp(44/T)$ | | | JPL 19-5 | Products unknown |

**Table S1.** Potential $NF_3$ atmospheric destruction pathways.

| Reactants | Products | Rate Law [cm3/molecule s] | Valid Temperature Range [K] | Rate at 298K | Source | Note |
|---|---|---|---|---|---|---|
| $SF_6 + H$ | $HF + SF_5$ | $3.32 \times 10^{-9} \exp(-15154.3/T)$ | 1460 - 1700 | N/A | NIST | High temp rxn |
| $SF_6 + CCl$ | Products | $2.41 \times 10^{-14}$ | 298 | $2.41 \times 10^{-14}$ | NIST | Products unknown |
| $SF_6 + CH$ | Products | $<5.0 \times 10^{-17}$ | 297 | N/A | NIST | Products unknown |
| $SF_6$ | $SF_5 + F$ | $3.38 \times 10^{17} (T/298)^{-1.90} \exp(-50754.9/T)$ | 1700 - 2000 | N/A | NIST | First-order reaction; Unit: [s-1] |
| $CF_3 + SF_6$ | $CF_4 + SF_5$ | $1.66 \times 10^{-13} \exp(-10100.5/T)$ | 303 - 638 | N/A | NIST | |
| $CH_3 + SF_6$ | $CH_3F + SF_5$ | $3.32 \times 10^{-11} \exp(-7099.7/T)$ | 413 - 443 | N/A | NIST | |
| $O(1D) + SF_6$ | $O(1D) + SF_6$ | $1.79 \times 10^{-14}$ | 298 | $1.79 \times 10^{-14}$ | NIST | |
| | | N/A | | $1.80 \times 10^{-14}$ | JPL 19-5 | |

**Table S2.** Potential $SF_6$ atmospheric destruction pathways.

| Atom | Formula | Name | Boiling Point | Phase (at STP) |
|------|---------|------|---------------|----------------|
| B | BF$_3$ | Boron Trifluoride | −100.3 | gas |
| B | B$_2$F$_4$ | Diboron Tetrafluoride | −34 | gas |
| Si | SiF$_4$ | Silicon Tetrafluoride | −90.3 | gas |
| Ge | GeF$_4$ | Germanium Tetrafluoride | −36.5 | gas |
| N | NF$_3$ | Nitrogen Trifluoride | −128.75 | gas |
| P | PF$_3$ | Phosphorus Trifluoride | −101.8 | gas |
| P | P$_2$F$_4$ | Diphosphorus Tetrafluoride | −6.2 | gas |
| As | AsF$_3$ | Arsenic Trifluoride | 57.8 | liquid |
| S | SF$_6$ | Sulfur Hexafluoride | −50.8 | gas |
| Se | SeF$_6$ | Selenium Hexafluoride | −34.5 | gas |

**Table S3.** Fully fluorinated molecules and their physical and chemical properties. STP is standard temperature and pressure.

## SI 4 Thermodynamics of Volcanic SF$_6$ and NF$_3$ Formation

We have estimated the amount of SiF$_4$, SF$_6$ and NF$_3$ that could be volcanically produced under a variety of atmospheric conditions. The conditions are those at which the gas erupts, and so is rapidly thermodynamically quenched into low temperature atmosphere, not necessarily conditions in the atmosphere. Below we show the analysis of all combinations of the following conditions:
- Temperature from 500K to 1500K in 100K steps
- 8 mineral redox buffers (QIF, IW, WM, IM, CoCoO, FMQ, NiNiO, MH)
- Abundance of water from 5% to 90% in 5% steps
- Pressure of 1, 10, 100, 1000, 10,000, 100,000 bar
- total gas sulfur content of 0.001, 0.01 or 0.1
- total gas HF content of 0.0001, 0.001, 0.01

The above conditions give a total of 85536 combinations. We assume that the 'activity' of $SiO_2$ is 1, and that $SiO_2$ is present as quartz: all other species are assumed to be gases. Abundances of $H_2$, $O_2$, $H_2S$ and $SO_2$ are calculated from the oxygen fugacity (set by the mineral redox buffer) and the abundance of water and total sulfur.

Plotting the amount of $SiF_4$ vs the amount of $SF_6$ or $NF_3$ shows that under almost all conditions the amount of $SF_6$ and $NF_3$ is less than $SiF_4$ by many orders of magnitude (Figure S1).

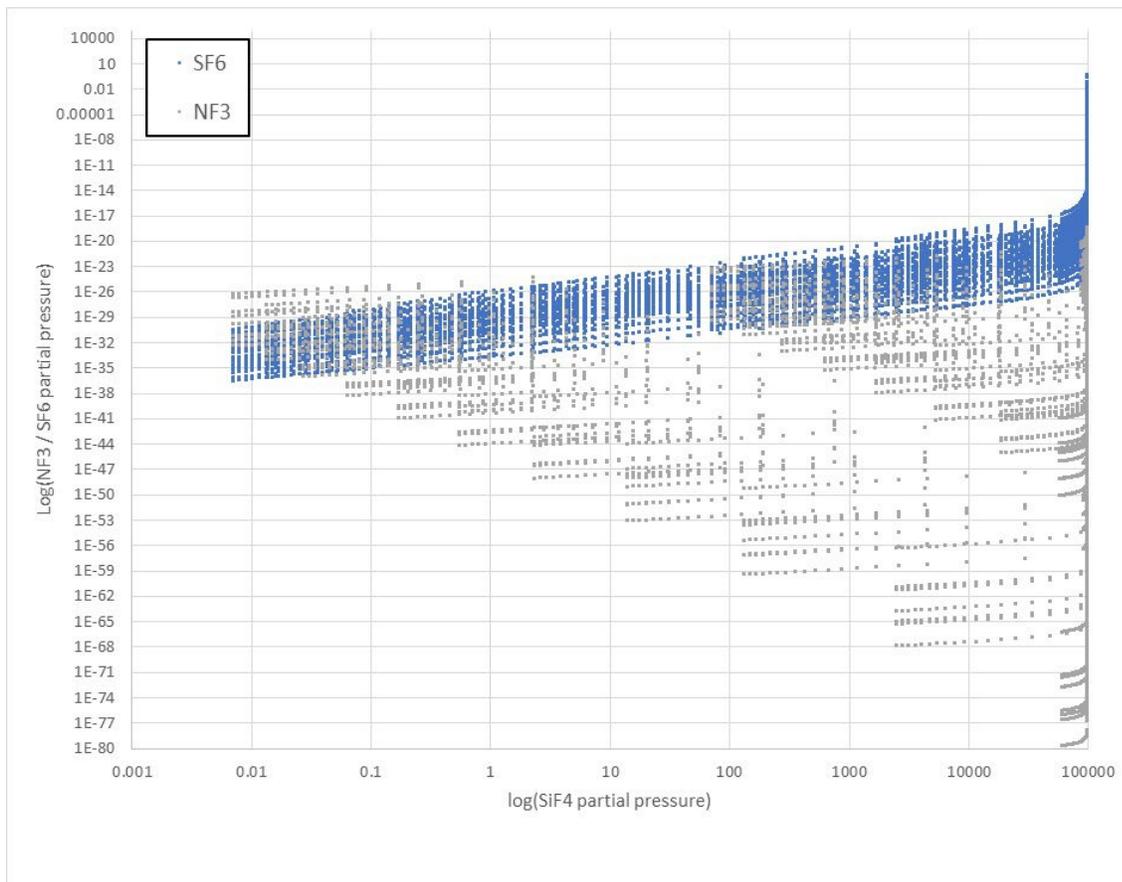

**Figure S1.** The amount of $SiF_4$ compared to the amounts of $NF_3$ and $SF_6$ produced by volcanoes under 85,536 combinations of tested conditions. x axis: partial pressure of $SiF_4$, y axis: partial pressure of $SF_6$ (blue dots) and $NF_3$ (grey dots). Under almost all conditions the amount of $SF_6$ and $NF_3$ is less than $SiF_4$ by many orders of magnitude.

In a few sets of conditions shown on the Figure S1 the $SF_6$ is a few % of $SiF_4$ (top right area on the Figure S1). Those conditions are high pressure, low water, low temperature conditions where thermodynamics predicts that the atmosphere is composed almost entirely of $SiF_4$, i.e. under those conditions the crustal rocks are being liquified and turned into $SiF_4$. Such a scenario is physically and stoichiometrically implausible. (We assume an infinite reservoir of F in the rocks, which is implausible, although not impossible given the mass of the crust compared to the mass of the atmosphere.). We have therefore plot the abundance ratios of $SF_6/SiF_4$ and $NF_3/SiF_4$ (Figure S2). Plotting the ratios of $SF_6/SiF_4$ and $NF_3/SiF_4$ allows for the partial pressure of $SiF_4$ to be arbitrary, therefore reflecting the ratios more accurately.

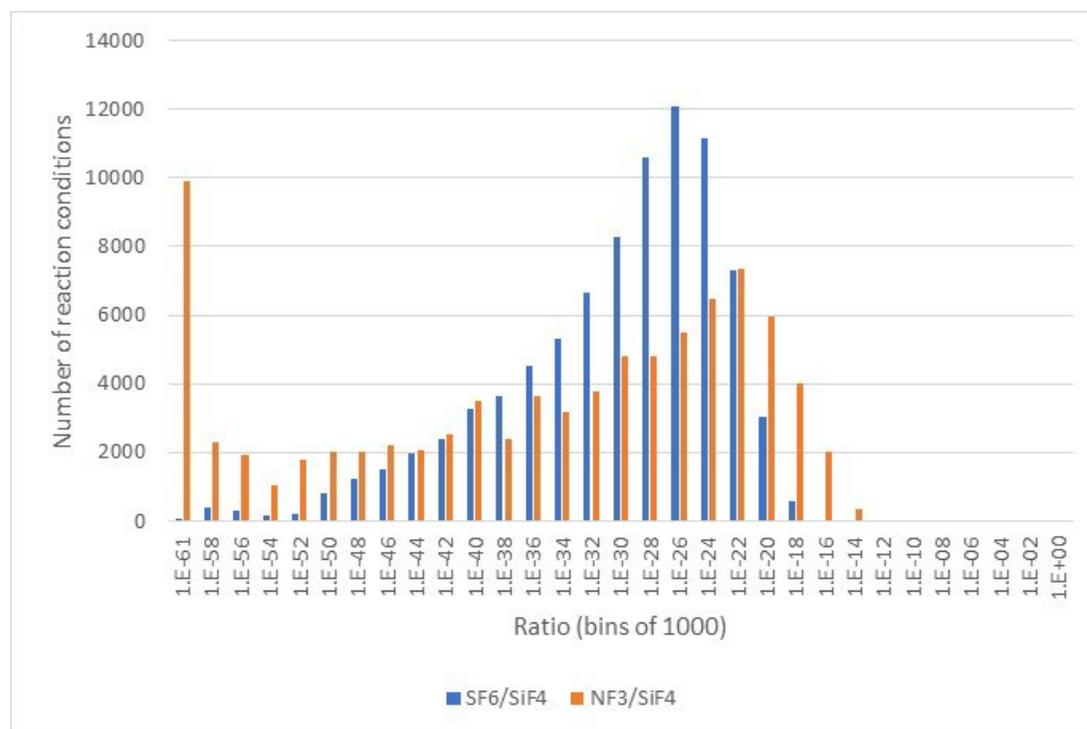

**Figure S2.** Abundance ratios of $SF_6/SiF_4$ and $NF_3/SiF_4$. y axis: number of reactions conditions. x axis: abundance ratios. No significant amounts of $NF_3$ and $SF_6$ are produced by volcanoes under 85,536 combinations of tested conditions.

## SI 5 Supplementary References